\documentclass[11pt]{article}
\pdfoutput=1
 \usepackage{mciteplus}
 \usepackage{tikz}
 \usepackage{color}
 \usepackage{xcolor}
 \usepackage{comment}
 \definecolor{darkblue}{rgb}{0.1,0.1,.7}
 \usepackage[colorlinks, linkcolor=darkblue, citecolor=darkblue, urlcolor=darkblue, linktocpage,hyperfootnotes=false]{hyperref} 
\usepackage{epsfig}
\usepackage{graphicx}
\usepackage{cite}
\usepackage{amsfonts}
\usepackage{amssymb}
\usepackage{bm}
\usepackage{latexsym}
\usepackage{mathtools}
\usepackage{amsmath}
\numberwithin{equation}{section}
\def\bq{\begin{quote}}
\def\eq{\end{quote}}
 at 10truept

\newcommand{\calc}{{\cal C}}

\newcommand{\calo}{{\cal O}}

\newcommand{\beq}{\begin{equation}}
\newcommand{\eeq}{\end{equation}}
\newcommand{\beqa}{\begin{eqnarray}}
\newcommand{\eeqa}{\end{eqnarray}}
\newcommand{\bea}{\begin{eqnarray}}
\newcommand{\eea}{\end{eqnarray}}

\newcommand{\hf}{\frac{1}{2}}

\def\roughly#1{\raise.3ex\hbox{$#1$\kern-.75em\lower1ex\hbox{$\sim$}}}

\def \bal#1\eal  {\begin{align} #1 \end{align}}
\def\({\left(}
\def\){\right)}
\def\[{\left[}
\def\]{\right]}
\def\<{\left\langle}
\def\>{\right\rangle}
\def\d{\mathrm{d}}

\newcommand{\nn} {\nonumber\\~}

\newcommand{\bra}{\langle}
\newcommand{\ket}{\rangle}

\newcommand{\sech}{{\rm sech}}
\newcommand{\ta}{{\tilde a}}

\begin{document}

\thispagestyle{empty}
\pagenumbering{Alph}
\begin{titlepage}
  \bigskip

  \bigskip\bigskip

  \bigskip

\begin{center}
{\Large \bf {Leading gravitational dressing of operators and states}\\ { in de Sitter space
}}
    \bigskip
\bigskip
\end{center}

  \begin{center}

 \rm {Steven B. Giddings\footnote{\texttt{giddings@ucsb.edu}}  and Zi-Yue Wang\footnote{\texttt{zi-yue@ucsb.edu}}}
  \bigskip \rm
\bigskip

{Department of Physics, University of California, Santa Barbara, CA 93106, USA}  \\
\rm

  \bigskip \rm
\bigskip
 
\rm

\bigskip
\bigskip

  \end{center}

\vspace{3cm}
  \begin{abstract}
  We describe construction of gravitationally dressed observables in de Sitter (dS) space, to leading non-trivial order in Newton's constant.  We show that for an underlying field theory observable to be ``gravitationally dressable," it must be dS invariant.  If it satisfies this condition, its gravitational dressing can be constructed in terms of certain Green functions, by solving the condition that the dressed observable commute with the constraints.  We provide a construction of the Green functions relevant for dressing observables lying on a time-symmetric slice.  A simple example of a dS invariant observable is an operator creating a two-particle state of scalars.  In the large mass limit, the particles are shown to be antipodally located, and the dressing of this operator (or corresponding state) can be explicitly constructed.  Its behavior is diagnosed by certain correlators involving this state, and we check explicitly that these correlators then reproduce a linearized version of the Schwarzschild-dS solution.  The underlying dS-invariant observables may be thought of as relational, and we argue can be generalized to include systems with properties of observers, which then could be likewise gravitationally dressed.  We also briefly describe how such dressed observables can be related to alpha vacua, and discuss other questions regarding a more complete construction and role of such operators.

 \medskip
  \noindent
  \end{abstract}

  \end{titlepage}
\pagenumbering{arabic}

\section{Introduction}

An important conceptual problem in quantum gravity is to understand its observables.  In fact, there are different things that are commonly meant by observables.  There are the things that observers observe, and more formally there are gauge-invariant operators that act on a Hilbert space of states.  In other familiar theories these are connected; part of the subtlety of quantum gravity is understanding and describing the connection.  For clearer terminology the latter might be referred to as q-observables.

This paper will focus on construction of observables in the sense of q-observables, bearing in mind that some connection to observations made by observers is expected.  The q-observables (with q- omitted henceforth) are anticipated to be closely related to the fundamental structure of the theory.  Specifically, if one takes the example of local quantum field theory, locality itself is characterized in terms of local subalgebras of observables\cite{Haag}.  And, interestingly, we have learned (see, {\it e.g.} \cite{Torr}) that gravity does not have local observables in the same sense.  A general class of nonlocal observables, with approximately local properties, in certain contexts, is constructed {\it relationally}; for some recent discussion of this see {\it e.g.} \cite{SGOO}.  

One type of relational observables is that of {\it gravitationally-dressed} observables, with concrete examples provided in \cite{DoGi1,DoGi2,DoGi3,GiKi,QGQF,DoGi4,GiPe2} and other works.\footnote{Earlier work connected to gravitational dressing includes \cite{Heem} and \cite{KaLigrav}; the first argued that the constraints lead to nontrivial commutators, but didn't give the dressed operators; the second focussed on constructing commuting operators.}  These are found by beginning with an observable of a field theory that is coupled to gravity, and adding a gravitational dressing to construct an operator that is gauge-invariant, specifically by solving the condition that it commutes with the constraints, which generate gauge transformations.  These are manifestly nonlocal, and as a result yield an interesting and apparently important modification of the algebraic structure of quantum field theory, that appears intimately connected to what is commonly thought of as holography\cite{Maroholo,SGholo}, and connects to other interesting and puzzling non-local structure of gravity\cite{SGOO}.\footnote{One also finds the well-noted transition from von Neumann algebras of type III to type II associated with the crossed product\cite{Wittcross,CPW} as a {\it truncation} of this structure\cite{SGcross}, also emphasizing the interest in the more general structure.}

One important role for observables is to change the state:  acting on one state, {\it e.g.} the vacuum, they produce another, different, state.  For example, if in field theory, an operator $O$ acts on the vacuum to create a particle or particles, a corresponding gravitationally-dressed observable $\hat O$ creates that particle or particles, dressed with an appropriate gravitational field.

Different examples have been constructed perturbatively in an expansion in $\kappa=\sqrt{32\pi G}$, where $G$ is Newton's constant. 
These include gravitational line-dressings, either in an otherwise flat or anti-de Sitter background\cite{DoGi1,GiKi}, or more symmetric dressings that, for example, produce the leading-order Schwarzschild solution\cite{DoGi1}, and a general prescription for constructing leading-order dressings about a general background has been described in \cite{GiPe2}.  A typical feature of such constructions, supported by a {\it dressing theorem}\cite{DoGi2,GiKi}, is that the gravitational dressing must extend to  asymptotic spatial infinity.

An important question, also connecting to interesting cosmological questions, has been that of construction of observables in closed cosmological spacetimes such as de Sitter space (dS), with no asymptotic spatial infinity.  This paper addresses this question, again to leading order in $\kappa$, using the general construction developed in \cite{GiPe2}.  An interesting aspect of this question is how such constructions avoid a dressing theorem.  Of course, a clue to this is that there are perfectly good solutions of the constraints of general relativity which describe massive excitations of dS, such as the Schwarzschild-dS solution; this latter case, of course, describes {\it two} massive bodies.  In fact, an analog is that of electromagnetism, where one cannot have a single charged particle in a compact space, but one can have two oppositely-charged particles.  

When approached using the general construction of \cite{GiPe2}, as we will describe in section two, one needs to find operators that commute with 
the constraints.  In order to consistently construct such operators at order $\kappa$, we show that the order $\kappa^0$ operator $O$ must satisfy an order $\kappa^0$ condition that it be dS invariant.  This is a current translation of older discussions of linearization instability\cite{FMM,DEHK}.  If this condition is satisfied, then gravitational dressings can be constructed by solving a consistent Green function problem.

Section three then describes construction of the Green functions solving this Green function problem, for the case of operators defined on a maximally-symmetric spatial slice of dS; these are constructed in terms of functions on the resulting $S^d$. 

Section four then gives a specific example.  An operator $O_2$ is constructed that creates two-particle states and satisfies the $\calo(\kappa^0)$ dS-invariance condition.  Applying the general dressing prescription then allows construction of a dressed operator $\hat O_2$.  To understand properties of these operators, we look at correlators where they act on the vacuum.  We show that, in a large mass limit, the operator $O_2$ creates particles at antipodal poles of $S^d$.  And, we concretely show that the operator $\hat O_2$ creates such particles, together with a gravitational field that matches, to leading order in $\kappa$, the Schwarzschild-dS solution.  

We close with some broader discussion of connections, implications, and open problems.  Specifically, our construction furnishes an example of a state with one particle relationally described with respect to another, which we argue could be generalized so that one of the two has properties like that of a more sophisticated observer, and thus give a way of implementing such an observer with pre-existing fields, rather than postulating observers in the fundamental description.  Such states should be equally-well dressed by our leading-order gravitational dressing construction.  Also, the various operators $\hat\calo_I$ act on the dS vacuum to create non-trivial states; describing multi-universe versions of these states appears to connect to discussions of $\alpha$ vacua, such as considered in \cite{Cole,GiSt}, revived in \cite{MaMa}, and explored in many other references.  Other questions include constructions on more general slicings, the question of time-dependence and definition of the Hilbert space\cite{GiPe3}, and the important question of higher-order or nonperturbative generalizations of these constructions.  The latter, in particular, are argued to have important implications for holographic behavior\cite{SGholo}, and for very novel algebraic structure of the theory called ``algebraic spacetime disruption\cite{SGOO}," associated with presumed quantum black hole behavior.

Two appendices summarize properties of functions such as harmonics on $S^d$, and explicit construction of some of the relevant Green functions.

\section{Leading gravitational dressing: general construction}
\label{dresssec}

We are interested in gauge invariant observables in general quantum spacetimes, both because of their connection to changes of the quantum state, and in connection to actual observation.  The general such observables are expected to be very complicated, but working perturbatively about a given background, we can find a simpler class of such observables that also have interesting physical interpretations.

The simpler observables we describe are constructed by starting with an observable of an underlying non-gravitational field theory, and then ``gravitationally dressing" that observable.  The observable we start with will not necessarily be invariant under the gauge symmetries of gravity, which in the geometric approximation are diffeomorphisms.  However, one can express the gauge-invariance condition order by order in the gravitational coupling, $\kappa$, and then seek a solution of it in such an expansion.  Following earlier work\cite{DoGi1,DoGi2,DoGi3,GiKi,QGQF,DoGi4,GiPe2}, this paper will seek such solutions at leading non-trivial order in $\kappa$.  

We will find these gravitationally-dressed observables by working perturbatively in a canonical description.  
Specifically, given a $D=d+1$-dimensional background, we can choose a time-slicing of that background, and corresponding ADM decomposition of the metric,
\beq
ds^2=-N^2 dt^2 +q_{ij}(dx^i+N^i dt)(dx^j+N^jdt)\ .
\eeq
Consider a time slice determined by a fixed $t$, and operators restricted to that slice.\footnote{Here we implicitly adopt a Heisenberg picture description of these operators.}
The constraints
\beq\label{cons}
\calc_\mu=\sqrt q \left( -\frac{G_{\mu\nu}}{8\pi G} + T_{\mu\nu} \right) n^\nu\ ,
\eeq
with $G_{\mu\nu}$ the Einstein tensor, $T_{\mu\nu}$ the stress tensor, and
\beq
n^\mu=\frac{1}{N} (1,-N^i)\ .
\eeq
the unit normal to the slices, generate the action of diffeomorphisms on these operators.  Specifically, consider the integral
\beq
C[\xi] =\int d^dx  \xi^\mu {\cal C}_\mu
\eeq
over the slice of fixed $t$.  For spatial vectors $\xi^\mu=(0,\xi^i)$, this will generate diffeomorphisms of the quantum fields and metric perturbations, and of their conjugate momenta, within the slice.  In treating the metric perturbations, we expand about the background,
\beq\label{metexp}
\tilde g_{\mu\nu} = g_{\mu\nu} + \kappa h_{\mu\nu}\ ,\ \tilde q_{ij}= q_{ij} +\kappa h_{ij} \ ;
\eeq
then the expansion of the constraints \eqref{cons},  expressed in terms of the metric $\tilde g_{\mu\nu}$, generate these transformations.
Likewise, if the equations of motion are satisfied, $C[\xi]$ with vector $\xi^\mu=Nn^\mu$ generates temporal diffeomorphisms.  These statements are explicitly checked for example in \cite{GiPe2}, for the case of a scalar field $\phi$ coupled to gravity.

Thus given a field operator $O$ defined at $t$, one may dress it with metric perturbations to perturbatively solve the conditions
\beq\label{invarc}
[\calc_\mu(t,x),\hat O(t)] =0\ 
\eeq
and this gives a gravitationally dressed, diffeomorphism-invariant version of $O$.

Such dressed operators were initially constructed for flat backgrounds in \cite{DoGi1},  have been studied further in \cite{DoGi2,DoGi3,GiKi,QGQF,DoGi4,GiPe2}, and have also been constructed for anti de Sitter backgrounds in \cite{GiKi}.  Acting on a state such as the vacuum, the dressed operator $\hat O$
can be interpreted as creating both the particles or fields associated with $O$, and also their corresponding gravitational fields.  The dressing is non-unique, corresponding to the fact that a given matter configuration can have different gravitational field configurations, differing for example by a gravitational wave configuration.  
Ref.~\cite{GiPe2} describes the extension of these constructions to more general backgrounds.  However, we expect additional subtleties in the case where space is  compact.  The present paper will focus in particular on perturbations of de Sitter space, in a global description.

Specifically, consider a compact time slice through de Sitter (dS) space.  As was discussed in \cite{DoGi1,DoGi2,DoGi3,GiKi,GiPe2}, we expect to be able to solve the invariance condition \eqref{invarc}, to leading order in  the $\kappa$ expansion, by an operator of the form
\beq\label{odress}
\hat O\simeq e^{i\int d^dx \sqrt q n^\nu V^\mu(x) T_{\mu\nu}} O e^{-i\int d^dx \sqrt q n^\nu V^\mu(x) T_{\mu\nu}}\ ,
\eeq
with integrals over the time slice, and with $V^\mu$ a certain functional of the metric perturbation $h_{ij}$ and its (densitized) conjugate $p^{ij}$, which satisfies the canonical commutators
\beq\label{ccrh}
[p^{ij}(t,x),h_{kl}(t,x')] = -\frac{i}{2}\left(\delta^i_k \delta^j_l+\delta^i_l \delta^j_k\right)\delta^d(x-x')\ .
\eeq
For the leading-order dressing, these functionals are postulated to be linear functionals of the form,
\beq\label{gend}
V^\mu(x)=\frac{\kappa}{2} \int d^{d}x' \left[H^\mu_{ij}(x',x) p^{ij}(x')+G^{ij\mu}(x',x) h_{ij}(x')\right]\ ,
\eeq
where  $H^\mu_{ij}$ and $G^{ij\mu}$ are c-number functions determined by the invariance condition \eqref{invarc}.  In this case the exponentials in  \eqref{odress} are Weyl operators acting on the metric perturbations.

To solve the invariance conditions \eqref{invarc} perturbatively in $\kappa$, we expand the constraints about the background metric $g_{\mu\nu}$ using \eqref{metexp}.  If we assume that $g_{\mu\nu}$ satisfies the constraints with vanishing matter fields, the constraints take the form\footnote{With non-zero matter backgrounds, there are additional terms given in \cite{GiPe2}.}\cite{GiPe2}
\beq
\tilde \calc_\mu(x) = \frac{1}{\kappa} L_\mu(h_{ij},p^{ij}) + \sqrt q (T_{n\mu} + t_{n\mu}) 
\eeq
where the leading $\calo(1/\kappa)$ terms $L_\mu$ are linear in the perturbations $h_{ij}$, $p^{ij}$, and where nonlinear terms in metric perturbations are collected in the expression $t_{n\mu}$, which can be interpreted as components of a stress-energy pseudotensor for the gravitational degrees of freedom.
Expanding  the invariance condition \eqref{invarc}  in $\kappa$ then gives
\beq\label{keqn}
0=[\tilde \calc_\mu(x),\hat O]= \frac{i}{\kappa} \int d^d x' [L_\mu(x),V^\nu(x')][\sqrt q T_{n\nu}(x'),O] + [\sqrt q T_{n\mu}(x),O] + \calo(\kappa)\ . 
\eeq

To leading order in $\kappa$, \eqref{keqn} then gives Green function equations for the functions  $G^{ij\mu}$ and $H^\mu_{ij}$, as described in \cite{GiPe2,SGcross}.  However, working with closed spatial sections like in de Sitter space introduces additional subtleties in finding the corresponding solutions.

Specifically, for the case of the vacuum de Sitter background, the metric has the Killing vectors $\xi^\mu_I$, and it can be shown that for the integral over such a closed surface in de Sitter \cite{DEHK},
\beq\label{Lint}
\int d^dx \xi^\mu_I L_\mu =0\ .
\eeq
Na\"\i vely we might have expected to solve \eqref{keqn} by requiring, as in the noncompact case,
\beq\label{Lcomm}
[\frac{1}{\kappa}L_\mu(x), V^\nu(x')]= i \delta_\mu^\nu \delta^d(x-x')\ .
\eeq
However, integrating this equation against $\xi^\mu_I$ would then give 
\beq
\int d^dx \xi^\mu_I [\frac{1}{\kappa}L_\mu(x), V^\nu(x')] =i \xi^\nu_I(x')\ ,
\eeq
which is inconsistent with \eqref{Lint}.  

The inconsistency can be avoided by introducing an additional ``background charge" term into \eqref{Lcomm}, so that it becomes
\beq\label{Lbackgroundcharge}
[\frac{1}{\kappa}L_\mu(x), V^\nu(x')]= i \delta_\mu^\nu \delta^d(x-x') - i c_\mu{}^\nu(x,x')\ ;
\eeq
the choice 
\beq\label{backgroundcharge}
c_\mu{}^\nu(x,x')= \sqrt{q(x)} \sum_{I,J} \(M^{-1}\)^{IJ} \xi_{I\mu}(x)\xi_J^\nu(x') \ ,
\eeq
with Graham matrix $M_{IJ}$ defined as 
\beq
M_{IJ}= \int \d^d x \, \sqrt{q(x)}\xi_I^\mu(x) \xi_{J\mu}(x) \ .
\eeq
is then consistent with \eqref{Lint}.  

Even with such a Green function, going back to  \eqref{keqn} and integrating it against  $\xi_I^\mu$ shows that the expansion is consistent 
with \eqref{Lint} if and only if
\beq\label{invcond}
\Big[\int d^dx\sqrt q  \xi^\mu_I(x) T_{n\mu}(x), O\Big] =0\ .
\eeq
This argument, which is a version of known  ``linearization instability" arguments\cite{FMM,DEHK}, in essence says that 
an operator $O$ must carry total charge zero with respect to the symmetries generated by the Killing vectors, that is, it must be invariant under the de Sitter (dS) isometries,  if it is to have a consistent gravitational dressing.

Our present goal will be to find dressings of operators satisfying the dS invariance conditions \eqref{invcond}.  To illustrate the basic construction in a simple example, we will focus on the case where the time slice is an extremal slice, with instantaneously vanishing extrinsic curvature.  This significantly simplifies the more general formulas in \cite{GiPe2}, and in particular the operators $L_\mu$ become
\beq\label{Lds}
L_n=-2\sqrt q\, L^{ij} h_{ij}\ ,\qquad 
L_i=-2\sqrt q\,q_{ij}D_k\left(\frac{p^{jk}}{\sqrt q}\right)\ .
\eeq
Here
\beq\label{Lijdef}
L^{ij} h_{ij}=D^iD^j h_{ij}-D^2 h-R_q^{ij}h_{ij} \ ,
\eeq
$h=q^{ij}h_{ij}$, $D^2=D_iD^i$, and all geometric quantities are those of the unit $S^d$ metric on the $\tau=0$ slice of global de Sitter,
\beq\label{globaldS}
ds^2=-d\tau^2+\cosh^2\tau\,d\Omega_d^2\ ;
\eeq
here we use units where the de Sitter radius $R=1$. 
Thus $N=1$, $N^i=0$, $K_{ij}=0$, and 
\beq\label{Riccieq}
R^q_{ij}=(d-1)q_{ij}
\eeq
on the slice.  With this time symmetric choice the mixed pieces of the dressing may be consistently set to zero, $H^k_{ij}=0$ and $G^{ij\tau}=0$, giving the simpler expressions
\beq\label{Vdsslice}
V^\tau(x)=\frac{\kappa}{2}\int d^d x'\,H_{ij}(x',x)p^{ij}(x')\ , \quad
V^k(x)=\frac{\kappa}{2}\int d^d x'\,G^{ijk}(x',x)h_{ij}(x')\ .
\eeq

In this simple case the equations \eqref{Lbackgroundcharge} for the dressings \eqref{Vdsslice} reduce to the Green function equations
\bal\label{dsgreeneqH}
-\sqrt{q(x)}\,L_x^{ij}H_{ij}(x,x')&=\delta^{d}(x-x')-c_\tau{}^\tau(x,x')\ ,\\
\label{dsgreeneqG}
q_{ij}(x)D^x_kG^{jkl}(x,x')&=\delta_i{}^l\,\delta^{d}(x-x')-c_i{}^l(x,x')\ .
\eal
Here the super/subscript $x$ indicates which argument the derivative acts on, and the background charges $c_\mu{}^\nu$ are those defined in \eqref{backgroundcharge}.  To construct the gravitational dressings, we must therefore solve these equations to find the corresponding Green functions.

\section{dS Green functions}
\label{gfsec}

We next solve the Green function equations \eqref{dsgreeneqH} and \eqref{dsgreeneqG} on the $\tau=0$ slice of global de Sitter.  Throughout this section $x,x'$ denote points on the unit sphere $S^d$, and
\beq
\label{udef}
u(x,x')=\cos\gamma(x,x')
\eeq
is the cosine of the geodesic distance between them.  Conventions and formulas for scalar and vector spherical harmonics are collected in Appendix~\ref{sphericalharmonics}.  In particular,
the scalar spherical harmonics $Y_{\ell a}$ satisfy
\beq\label{scalarharmonics}
D^2Y_{\ell a}=-\ell(\ell+d-1)Y_{\ell a}\ ,\qquad
\sum_{\ell=0}^{\infty}\sum_a Y_{\ell a}(x)Y_{\ell a}(x')=\frac{\delta^{d}(x-x')}{\sqrt {q(x)}}\ .
\eeq

We will find  the solutions in terms of scalar Green functions $G_s^{d}(u)$ satisfying the equation
\beq\label{Sgreen}
\frac{1}{\operatorname{Vol}(S^d)}\left[D^2+s(s+d-1)\right]G_s^{d}(u)
=\frac{\delta^{d}(x-x')}{\sqrt {q(x)}}
-\sum_a Y_{sa}(x)Y_{sa}(x')\ ,
\eeq
using \eqref{udef}, for chosen integer $s$.
The subtraction of the $\ell=s$ harmonics is required because these modes are annihilated by the operator on the left.  Fixing $G_s^d$ to have no $\ell=s$ component, we then find the solutions (see Appendix~\ref{gfapp})
\beq\label{Gsdef}
G_s^{d}(u)=
\sum_{\ell=0,\,\neq s}^{\infty}
\frac{(2\ell+d-1)\,C_\ell^{\frac{d-1}{2}}(u)}{(d-1)(s-\ell)(\ell+d+s-1)}\ ,
\eeq
where $C_\ell^{\frac{d-1}{2}}$ are Gegenbauer polynomials.  Examples of closed form expressions for $G_s^{d}$  are provided in 
Appendix \ref{gfapp}; this type of Green function was also recently used in \cite{MNX}.

\subsection{$H_{ij}$}

To solve \eqref{dsgreeneqH}, we use the Ansatz
\beq\label{Hans}
H_{ij}(x,x')=q_{ij}(x)\,{\cal H}(u)\ .
\eeq
Using the definition \eqref{Lijdef} and the Ricci tensor \eqref{Riccieq} we find
\beq\label{lijred}
L_x^{ij}\big(q_{ij}{\cal H}\big)=-(d-1)(D_x^2+d){\cal H}\ .
\eeq

In \eqref{dsgreeneqH}, we need $c_\tau^\tau$ which is given in terms of the Killing vectors by  \eqref{backgroundcharge}.  Explicit formulas for the Killing vectors can be found in \cite{Gibbons:2021zqu}.  The rotation Killing vectors are given by
\beq\label{rotvecs}
\xi^0_{R\ta}=0\ ,\quad \xi^i_{R\ta} = Y^{Ti}_{1\ta}\ ,
\eeq
where $Y^{Ti}_{\ell \ta}$ are the transverse vector harmonics (see Appendix \ref{sphericalharmonics}), and the boost Killing vectors are given in terms of the $\ell=1$ scalar harmonics by
\beq
\xi^0_{Ba}= Y_{1a}\ , \quad \xi^i_{Ba}=\tanh\tau D^i  Y_{1a}\ .
\eeq
Only the latter contribute to $c_\tau^\tau$, which becomes\footnote{In this basis, the Gram matrix is block diagonal with respect to the boost and rotation sectors. Working on the $\tau=0$ slice, we find its nonvanishing blocks
\bal\nonumber
(M_B)_{ab} = \int_{S^d} \d^d x\,\sqrt{q}\, g^{\tau\tau}\xi_{Ba,\tau}\xi_{Bb,\tau}=-\delta_{ab}\ ,\ (M_R)_{\ta\tilde b}=\int_{S^d} \d^d x\,\sqrt{q}\,g^{ij}\xi_{R\ta,i}\xi_{R\tilde b,j}=\delta_{\ta\tilde b}\ .
\eal
}
\beq
\frac{c_\tau{}^\tau(x,x')}{\sqrt {q(x)}}
=\sum_aY_{1a}(x)Y_{1a}(x')
=\frac{d+1}{\operatorname{Vol}(S^d)}\,u(x,x')\ ,
\eeq
where the second equality follows from the addition theorem, \eqref{additiontheorem}, and $C_1^\alpha(u)=2\alpha u$.  Combining this with equations \eqref{dsgreeneqH} and \eqref{lijred} gives the equation for $\cal H$,
\beq
(d-1)(D_x^2+d){\cal H}(u)
=\frac{\delta^{d}(x-x')}{\sqrt {q(x)}}
-\sum_aY_{1a}(x)Y_{1a}(x')\, ;
\eeq
this is \eqref{Sgreen} with $s=1$, and the solution with no $\ell=1$ component is
\beq\label{Hsolution}
{\cal H}(u)=\frac{1}{(d-1)\operatorname{Vol}(S^d)}\,G_1^{d}(u)\ .
\eeq
This and \eqref{Hans} provide an explicit solution $H_{ij}(x,x')$.

\subsection{$G^{ijk}$}

We next solve for the Green functions $G^{ijk}$ satisfying \eqref{dsgreeneqG}.  
Its source term  can be decomposed into longitudinal and transverse vector harmonics; for further discussion of their properties see  Appendix~\ref{sphericalharmonics}.  The longitudinal vector harmonics are given in terms of scalar harmonics by
\beq
Y^L_{i,\ell a}=\frac{1}{\sqrt{\ell(\ell+d-1)}}D_iY_{\ell a}\ ,\qquad \ell\geq 1\ .
\eeq
Formulas for the transverse harmonics are more complicated; these obey
\beq
D^iY^T_{i,\ell \ta}=0\ ,\qquad
D^2Y^T_{i,\ell \ta}=\left[-\ell(\ell+d-1)+1\right]Y^T_{i,\ell \ta}\ .
\eeq
The $\ell=1$ transverse modes give  the Killing vectors on $S^d$, as seen in \eqref{rotvecs}.  Therefore, for a real orthonormal basis of transverse harmonics, the background charge \eqref{backgroundcharge} is
\beq\label{cijvector}
\frac{c_i{}^j(x,x')}{\sqrt{{q}(x)}}=\sum_{\ta=1}^{d(d+1)/2}Y^T_{i,1\ta}(x) Y^{Tj}_{1\ta}(x')\,  \ .
\eeq
Using the completeness relation for vector harmonics, \eqref{vectorcomplete},
the source in \eqref{dsgreeneqG} then becomes
\beq\label{DGsource}
\frac{\delta^j_{i}\delta^{d}(x-x')-c_i{}^j(x,x')}{\sqrt {q(x)}}
=\sum_{\ell=1}^\infty\sum_aY^L_{i,\ell a}(x)Y^{Lj}_{\ell a}(x')
+\sum_{\ell=2}^\infty\sum_\ta Y^T_{i,\ell \ta}(x)Y^{Tj}_{\ell \ta}(x')\ .
\eeq

To solve  \eqref{dsgreeneqG}, we  try an  Ansatz of the form
\beq\label{Gansatz}
G^{ijk}(x,x')=\sqrt {q(x)}\left[q^{ij}(x){\cal T}^k(x,x')
+2D_x^{(i}{\cal G}^{j)k}(x,x')\right]\ ,
\eeq
where
\beq
D^x_j{\cal G}^{jk}=0\ .
\eeq
Then
\beq\label{DGsep}
\frac{1}{\sqrt{ q(x)}}D^x_jG_i{}^{jk}(x,x')
=D^x_i{\cal T}^k(x,x')+\left(D_x^2+d-1\right){\cal G}_i{}^k(x,x')\ ,
\eeq
where in the commutator of derivatives we have used the 
formula \eqref{Riccieq} for the Ricci tensor of $S^d$.

The Ansatz has thus separated the left hand side of \eqref{dsgreeneqG}
 into longitudinal and transverse pieces.  Comparing these pieces to the source \eqref{DGsource} on the right hand side then gives
the full solution
\bal\label{Tsolution}
{\cal T}_i(x,x')&=
\sum_{\ell=1}^{\infty}\sum_a\frac{1}{\sqrt{\ell(\ell+d-1)}}Y_{\ell a}(x)Y^L_{i,\ell a}(x')
=-\frac{1}{\operatorname{Vol}(S^d)}D_i^{x'}G_0^{d}(u)\ ,\\
\label{Gsolution}{\cal G}_i{}^j(x,x')&=
-\sum_{\ell=2}^{\infty}\sum_\ta
\frac{1}{(\ell-1)(\ell+d)}
Y^T_{i,\ell \ta}(x)Y^{Tj}_{\ell \ta}(x')\ , 
\eal
where we have related the longitudinal sum to the scalar Green function given by \eqref{Gsdef}, using the addition theorem \eqref{additiontheorem}.  
Closed form expressions for the transverse functions \eqref{Gsolution} can also be found.

\section{Example: bilinear operator and match to Schwarzschild-de Sitter}

The expressions \eqref{Hans}, \eqref{Hsolution} and \eqref{Gansatz}, together with the definitions \eqref{Vdsslice}, give the leading de Sitter dressing of a general operator $O$ satisfying the invariance condition \eqref{invcond}, on the extremal slice.  One approach to constructing such invariant operators is to begin with a non-invariant operator $\bar O$ and average it\cite{GiMa,KMYZ} over the de Sitter group $SO(D,1)$,
\beq\label{dsav}
O= \int d\Lambda U^\dagger (\Lambda) \bar O U(\Lambda) \ ,
\eeq
where $U(\Lambda)$ represents the general $SO(D,1)$ transformation $\Lambda$ acting on the fields.

In this section we will investigate a simple case of a bilinear operator, which creates a two-particle state.  We will find that the resulting dressed operator creates the linearized version of the Schwarzschild-dS solution, in the large mass limit.

\subsection{Invariant state and operator}

We begin by constructing the undressed invariant two-particle state, and then the corresponding invariant operator.
We work with a free massive scalar field with mass  $M^2=\mu^2+\frac{d^2}{4}$. This has the mode expansion\cite{Marolf:2008hg, Marolf:2012kh}
\bal
\label{modeexp}
\phi(\tau,x)=\sum_{\ell=0}^\infty\sum_{a=1}^{N_\ell}
\left[u_{\ell a}(\tau,x)a_{\ell a}
+u_{\ell a}(\tau,x)^*a_{\ell a}^\dagger\right]\, 
\eal
with mode functions 
\bal
u_{\ell a}(\tau,x)=f_\ell(\tau)Y_{\ell a}(x)\ .
\eal
Again, spherical harmonic conventions are summarized in Appendix \ref{sphericalharmonics}, and the normalized temporal wavefunctions are\cite{Marolf:2012kh}
\bal\label{temporalwavefunction}
f_\ell(\tau)&=2^{-\ell-\frac{d}{2}}
\frac{\sqrt{\Gamma(\ell+\frac{d}{2}+i \mu)
\Gamma(\ell+\frac{d}{2}-i \mu)}}{\Gamma(\ell+\frac{d+1}{2})} \nn
&\quad \times
(\cosh \tau)^\ell \ {}_2F_1\!\left(
 \ell+\frac{d}{2}-i\mu,\,
 \ell+\frac{d}{2}+i\mu;\,
 \ell+\frac{d+1}{2};\,
 \frac{1-i\sinh \tau}{2}
 \right)\ .
\eal

When describing the effects of boosts, it is useful to choose  spherical harmonics adapted to a chosen boost axis \cite{Marolf:2008hg}.
If we write the spatial metric as
\beq\label{metdecomp}
d\Omega_d^2=d\chi^2+\sin^2\chi\,d\Omega_{d-1}^2\ ,
\eeq
a  boost-adapted harmonic basis may be denoted by
$Y_{\ell k\alpha}^{d}$, where
\beq
\label{realb}
Y_{\ell k\alpha}^{d}(\chi,\Omega_{d-1})
={\cal N}_{\ell k}(\sin\chi)^k
C_{\ell-k}^{\,k+\frac{d-1}{2}}(\cos\chi)
Y_{k\alpha}^{d-1}(\Omega_{d-1})\ ,
\qquad 0\leq k\leq \ell\ .
\eeq
Here 
\beq
{\cal N}_{\ell k}=\frac{\Gamma(2k+d-1)}{2^{k+(d-1)/2}\Gamma(k+d/2)}\(\frac{(2\ell+d-1)\Gamma(\ell-k+1)}{\Gamma(\ell+k+d-1)}\)^{\frac{1}{2}}
\eeq
 is a normalization factor,  $\ell$ is the total
angular momentum on $S^d$, $k$ is the total angular momentum on the transverse
$S^{d-1}$, $C_p^\lambda$ is a Gegenbauer polynomial, and $\alpha$ labels the degeneracy of degree-$k$ harmonics
on $S^{d-1}$.

In a real basis \eqref{realb}, the states 
\beq\label{rotsing}
|S_\ell\ket=
\sum_{k=0}^{\ell}\sum_{\alpha}
a^\dagger_{\ell k\alpha}a^\dagger_{\ell k\alpha}|0\ket\ .
\eeq
are 
 singlets under general rotations on $S^{d}$.
To find invariant states under the full set of dS isometries, we consider a state of the form
\beq\label{gen2p}
|2\text{-particle}\ket
=\sum_{\ell=0}^{\infty}q_\ell |S_\ell\ket\ ,
\eeq
and solve the boost-invariance condition.

Consider a  boost associated to the decomposition \eqref{metdecomp}, generated by the Killing vector
\beq\label{specb}
B=\cos\chi\,\partial_\tau-\tanh\tau\,\sin\chi\,\partial_\chi\ .
\eeq
Following \cite{Marolf:2008hg}, its action on the wavefunction is 
\bal
&i{\cal L}_{B}u_{\ell k\alpha}
=\beta_{\ell k}\,u_{\ell+1,k\alpha}
+\beta_{\ell-1,k}u_{\ell-1,k\alpha}\ , \\
&\beta_{\ell k}
=\left\{
\frac{(\ell+k+d-1)(\ell-k+1)
\left[\left(\ell+\frac {d}{2}\right)^2+\mu^2\right]}
{(2\ell+d-1)(2\ell+d+1)}
\right\}^{1/2}\ .
\eal

To find the action on the two-particle states \eqref{gen2p}, let $\widehat B$ be the corresponding Hermitian boost operator, defined by
\beq
[\widehat B,\phi]=i{\cal L}_B\phi\ .
\eeq
Using the oscillator expansion of $\phi$, this gives the hermitian operator\cite{Marolf:2008hg}
\beq
\widehat B
=-\sum_{\ell k\alpha} a^\dagger_{\ell k\alpha}
\left(\beta_{\ell-1,k}\,a_{\ell-1,k\alpha}
+\beta_{\ell k}\,a_{\ell+1,k\alpha}\right)\ ,
\eeq
and hence
\beq
\widehat B\,a^\dagger_{\ell k\alpha}|0\ket
=-\left(\beta_{\ell k}\,a^\dagger_{\ell+1,k\alpha} +
\beta_{\ell-1,k}\,a^\dagger_{\ell-1,k\alpha}\right) |0\ket\ .
\eeq
Acting on the fixed-$\ell$ rotation singlet \eqref{rotsing} gives
\beq
\widehat B|S_\ell\ket
=\sum_{k\alpha}\widehat B
a^\dagger_{\ell k\alpha}a^\dagger_{\ell k\alpha}|0\ket 
=-2\sum_{k\alpha}\left(\beta_{\ell k}\,
a^\dagger_{\ell+1,k\alpha}a^\dagger_{\ell k\alpha}
+\beta_{\ell-1,k}\,
a^\dagger_{\ell-1,k\alpha}a^\dagger_{\ell k\alpha}\right)|0\ket\ .
\eeq
Boost invariance is the condition 
\beq
\widehat B \sum_{\ell=0}^\infty q_\ell |S_\ell\ket =0\ .
\eeq
Comparing the coefficient of each
$a^\dagger_{\ell+1,k\alpha}a^\dagger_{\ell k\alpha}|0\ket$ gives
\beq
q_{\ell+1}=-q_\ell\ ,
\qquad \Rightarrow \qquad
q_\ell=(-1)^\ell q_0\ , 
\eeq
and thus, up to normalization, taking $q_0=1$,
\beq\label{2pstate}
|2\text{-particle}\ket
=\sum_{\ell=0}^{\infty}(-1)^\ell |S_\ell\ket\ ;
\eeq
this generalizes the $d=1$ example given in \cite{KoLi}.
Since this state is rotation invariant, and an arbitrary boost can be converted to the boost \eqref{specb} by a rotation, this unique state is invariant under the full set of dS isometries.

We might expect the corresponding  invariant hermitian bilinear operator to be equally simple, which we now write using a general real basis for spherical harmonics
\bal\label{o2def}
O_2
=\sum_{\ell=0}^\infty (-1)^\ell\sum_{a=1}^{N_\ell}
\left(a_{\ell a}a_{\ell a}
+a_{\ell a}^\dagger a_{\ell a}^\dagger\right)\ .
\eal
Using equations analogous to those above, one can verify that $O_2$ indeed commutes with all dS generators.

\subsection{Dressing and operator diagnostics}

Given the simple example of the dS-invariant operator $O_2$, we can consider its dressed version, and explore the interpretation of the resulting operator.  The leading order dressed operator is, as in \eqref{odress},
\beq\label{o2dress}
\hat O_2 \simeq e^{i\int d^dx \sqrt q n^\nu V^\mu(x) T_{\mu\nu}} O_2 e^{-i\int d^dx \sqrt q n^\nu V^\mu(x) T_{\mu\nu}}\ ,
\eeq
with dressings $V^\tau$, $V^k$ given by \eqref{Vdsslice}, and where the Green functions $H_{ij}$ and $G^{ijk}$ are those found in Sec.~\ref{gfsec}.  One way to explore the physical interpretation of this operator is to consider the state it creates acting on the vacuum state.

We begin by investigating the undressed ({\it i.e} ${\cal O}(\kappa^0)$) version of this state, \eqref{2pstate}, or $O_2|0\rangle$. Properties of this state can be diagnosed through the correlation function $\bra0|\phi(0,x_1)\phi(0,x_2)O_2|0\ket$, which, using the mode expansion \eqref{modeexp} and the addition theorem \eqref{additiontheorem}, becomes
\bal\label{2ptdiag}
\bra0|\phi(0,x_1)&\phi(0,x_2)O_2|0\ket \nn
&=2\sum_{\ell=0}^\infty (-1)^\ell f_\ell^2(0)
\sum_{a=1}^{N_\ell}Y_{\ell a}(x_1)Y_{\ell a}(x_2) \nn
&= 2\sum_{\ell=0}^\infty (-1)^\ell f_\ell^2(0)
\frac{N_\ell}{\operatorname{Vol}(S^d)}
\frac{C_\ell^{\frac{d-1}{2}}\left[u(x_1,x_2)\right]}{C_\ell^{\frac{d-1}{2}}(1)} \ .
\eal
 We can see that this correlator is equal to twice the Bunch-Davies (vacuum) Wightman two-point function, evaluated with one of its arguments replaced by its spatial antipode.  
The Wightman two-point function is\cite{Einhorn:2002nu}
\bal
\bra 0|\phi(\tau_1,x_1)\phi(\tau_2,x_2)|0\ket &= \sum_{\ell=0}^\infty \sum_{a=1}^{N_\ell} f_\ell(\tau_1) f_\ell ^*(\tau_2) Y_{\ell a}(x_1) Y_{\ell a}(x_2)= 
\sum_{\ell=0}^\infty  f_\ell(\tau_1)f_\ell^*(\tau_2)
\frac{N_\ell}{\operatorname{Vol}(S^d)}
\frac{C_\ell^{\frac{d-1}{2}}\left[u(x_1,x_2)\right]}{C_\ell^{\frac{d-1}{2}}(1)} \nonumber\\
&=\frac{\Gamma\left(\frac{d}{2}+i\mu\right)
\Gamma\left(\frac{d}{2}-i\mu\right)}{(4\pi)^{\frac{d+1}{2}}
\Gamma\left(\frac{d+1}{2}\right)}{}_2F_1\left(\frac{d}{2}+i\mu,\frac{d}{2}-i\mu;\frac{d+1}{2};\frac{1+Z_{12}}{2}\right)\ ,
\eal
where
\beq
Z_{12}=-\sinh \tau_1 \sinh \tau_2 + \cosh \tau_1 \cosh \tau_2 u(x_1,x_2) \ .
\eeq
Setting $\tau_1=\tau_2=0$, one finds $f^*_\ell(0)=f_\ell(0)$ from \eqref{temporalwavefunction}.  Then the action of the antipodal transformation $A$ gives $u(x_1,Ax_2)=-u(x_1,x_2)$, and the equality is established using $C_\ell^\lambda(-u)= (-1)^\ell C_\ell^\lambda(u)$. 
The interpretation of this  is even clearer if one works in the large-mass limit, where
\beq\label{2ptlargem}
 f_\ell(0)^2
=\frac{1}{2\mu}+{\cal O}(\mu^{-3})\ .
\eeq
Then, \eqref{2ptdiag} becomes
\bal\label{2ptapprox}
\bra0|\phi(0,x_1)\phi(0,x_2)O_2|0\ket
&\simeq \frac{1}{\mu}
\sum_{\ell=0}^\infty \frac{N_\ell}{\operatorname{Vol}(S^d)}
\frac{(-1)^\ell C_\ell^{\frac{d-1}{2}}\left[u(x_1,x_2)\right]}{C_\ell^{\frac{d-1}{2}}(1)} \nn
&= \frac{1}{\mu}
\sum_{\ell=0}^\infty \frac{N_\ell}{\operatorname{Vol}(S^d)}
\frac{C_\ell^{\frac{d-1}{2}}\left[u(x_1,Ax_2)\right]}{C_\ell^{\frac{d-1}{2}}(1)}= \frac{1}{\mu}
\frac{\delta^{d}(x_1-A x_2)}{\sqrt {q(x_1)}}\ ,
\eal
where the last equality follows from completeness \eqref{compA} and the addition theorem \eqref{additiontheorem}.

The expressions \eqref{2ptdiag}, \eqref{2ptapprox} show that in the invariant state $O_2|0\rangle$, if a particle is found (annihilated) on the $\tau=0$ slice at $x_2$, another particle is located (annihilated) at the antipodal point, with precise localization in the large mass limit.

To diagnose behavior of the dressed state $\hat O_2|0\rangle$, we instead consider the correlator
\beq
\bra0|\phi(0,x_1)\phi(0,x_2)h_{ij}(0,x_3)\hat{O}_2|0\ket\ .
\eeq
While this means we are using a non-diffeomorphism invariant correlator to explore properties of the diffeomorphism-invariant state (due to the specified arguments $x_i$), we anticipate more complicated diffeomorphism-invariant correlators could perform similar diagnoses.

Using \eqref{o2dress}, the contribution of the leading dressing to this correlator is given by
\bal
&\bra0|\phi(0,x_1)\phi(0,x_2)h_{ij}(0,x_3)\hat{O}_2|0\ket \nn
&=i\int \d^d x\,
\bra0|\phi(0,x_1)\phi(0,x_2)h_{ij}(0,x_3)V^\nu(x)
[\sqrt q T_{n\nu}(x),O_2]|0\ket
+{\cal O}(\kappa^2)\ .
\eal
Here, and in the rest of this section, we work in the large $\mu$ limit; this implies that the  $V^iT_{ni}$ term is subleading since $T_{ni}$ contains only one time derivative, while the leading part of $T_{nn}$ contains two powers of the large frequency. 
On the time-reflection-symmetric slice,
$\bra0|h_{ij}(0,x)p^{kl}(0,x')|0\ket=\hf \langle0|[h_{ij}(0,x),p^{kl}(0,x')]|0\rangle$.  The canonical commutator \eqref{ccrh}, together
with \eqref{Vdsslice}, therefore give the large-$\mu$ expression
\bal\label{dressdiag}
&\bra0|\phi(0,x_1)\phi(0,x_2)h_{ij}(0,x_3)\hat{O}_2|0\ket \nn
&\simeq -\frac{\kappa}{4}\int \d^d x\,H_{ij}(x_3,x)\,
\bra0|\phi(0,x_1)\phi(0,x_2)
[\sqrt q T_{nn}(x),O_2]|0\ket
+{\cal O}(\kappa^2)\ .
\eal

Keeping only this leading part in large $\mu$, and using the large-$\mu$ expressions $\partial_\tau f_\ell(0)\simeq -i\mu f_\ell(0)$ and
$f_\ell(0)\simeq 1/\sqrt{2\mu}$ from \eqref{2ptlargem}, gives 
\bal
\sqrt q T_{nn}(x)
&=\frac{\sqrt q}{2}\left[
(\partial_\tau\phi)^2
+q^{ij}\partial_i\phi\partial_j\phi
+M^2\phi^2\right] \nn
&\simeq  \sqrt q \sum_{\ell a}\sum_{\ell' a'} \left[\partial_\tau f_\ell^*(0) \partial_\tau f_{\ell'}(0) + M^2 f_\ell^*(0)f_{\ell'}(0)\right]
Y_{\ell a}(x)Y_{\ell' a'}(x)
a_{\ell a}^\dagger a_{\ell' a'}\ ,\nn
&\simeq \mu\sqrt q
\sum_{\ell a}\sum_{\ell' a'}
Y_{\ell a}(x)Y_{\ell' a'}(x)
a_{\ell a}^\dagger a_{\ell' a'}\ ,
\eal
up to a normal-ordering constant, set to zero.
Here  the spatial derivative terms are subleading, and in the  $aa$  and $a^\dagger a^\dagger$ terms, the leading $\mathcal{O}(\mu)$ contributions from $(\partial_\tau\phi)^2$ and $M^2\phi^2$ cancel.

Write $\phi^\pm$ for the positive- and negative-frequency parts of the scalar field on the $\tau=0$ slice, so that $\phi^+$ contains only annihilation operators and $\phi^-$ contains only creation operators.  
In \eqref{dressdiag}, only the annihilation parts of the two external scalar fields contribute against $\bra0|$, so
\bal\label{tcorr}
&\bra0|\phi(0,x_1)\phi(0,x_2)
[\sqrt q T_{nn}(x),O_2]|0\ket \nn
&=
\bra0|\phi^+(0,x_1)\phi^+(0,x_2)
\sqrt q T_{nn}(x)O_2|0\ket \nn
&\simeq
\mu\left[\delta^{d}(x-x_1)+\delta^{d}(x-x_2)\right]
\bra0|\phi(0,x_1)\phi(0,x_2)O_2|0\ket\ ,
\eal
where we use the commutator
\beq
[\phi^+(0,x),\sqrt q T_{nn}(x')]
=\mu\delta^{d}(x-x')\phi^+(0,x)+{\cal O}(\mu^0)\ .
\eeq
This relates the expression to the earlier correlator \eqref{2ptdiag}, \eqref{2ptapprox}.

Combining eqs.~\eqref{dressdiag}, \eqref{tcorr}, and \eqref{2ptapprox} then gives the diagnostic correlator
\bal\label{3ptcorr}
&\bra0|\phi(0,x_1)\phi(0,x_2)h_{ij}(0,x_3)\hat{O}_2|0\ket \nn
&\simeq
-\frac{\kappa\mu}{4}
\left[H_{ij}(x_3,x_1)+H_{ij}(x_3,x_2)\right]
\bra0|\phi(0,x_1)\phi(0,x_2)O_2|0\ket \nn
&\simeq
-\frac{\kappa}{4}
\left[H_{ij}(x_3,x_1)+H_{ij}(x_3,A x_1)\right]
\frac{\delta^{d}(x_1-A x_2)}{\sqrt {q(x_1)}}\ .
\eal
This has the interpretation that the particles are antipodally located, as described above, and have a gravitational field given by the coefficient in \eqref{3ptcorr}.  In the next section, we will see that this field precisely matches that of the linearized Schwarzschild-dS solution.

\subsection{Linearized Schwarzschild-dS}

The $(d+1)$-dimensional Schwarzschild-dS metric can be written as
\beq\label{dSS}
\d s^2=-f(r)\d t^2+f(r)^{-1}\d r^2+r^2\d\Omega_{d-1}^2\ ,
\qquad
f(r)=1-r^2-\frac{2m}{r^{d-2}}\ .
\eeq
We normalize the parameter $m$ in terms of the physical mass $M$ by
\beq\label{Mrelat}
m=\frac{8\pi GM}{(d-1)\operatorname{Vol}(S^{d-1})}\ .
\eeq
For small $m$, the cosmological horizon radius and the dimensionless surface gravity are
\beq
r_c=1-m+{\cal O}(m^2)\ ,
\qquad
\alpha=\frac{|f'(r_c)|}{2}=1-(d-1)m+{\cal O}(m^2)\ .
\eeq
To extend the metric across the cosmological horizon, we introduce Kruskal coordinates \cite{Bazanski:1986, Guven:1990ubi}:
\beq\label{Kruskal}
x^\pm=t\pm r_*\ ,
\qquad
r_*=\int^r \frac{\d\rho}{f(\rho)}\ ,
\qquad
U=e^{\alpha x^-}\ ,
\qquad
V=-e^{-\alpha x^+}\ .
\eeq
In these Kruskal coordinates, the metric becomes
\beq
\d s^2=\frac{f(r)}{\alpha^2UV}\d U\d V+r^2\d\Omega_{d-1}^2\ .
\eeq

To extract the perturbation on the time-reflection-symmetric slice, however, we can work directly with the induced metric in the original coordinates, without using the Kruskal form of the metric. Restricting to the time-reflection-symmetric slice described by $t=0$, the spatial metric is 
\beq
\d s^2=f(r)^{-1}\d r^2+r^2\d\Omega_{d-1}^2\ .
\eeq
We now expand this metric about pure de Sitter to first order in $m$.
Consider the coordinate transformation $r=r_c\sin\chi=[1-m+\mathcal O(m^2)]\sin\chi$. At zeroth order in $m$, this gives the round-sphere metric induced on the $\tau=0$ slice of \eqref{globaldS}. The leading perturbation from \eqref{dSS} is
\beq\label{SdSslicepert}
\delta q_{ij}\d x^i\d x^j
=2m\left[
\frac{\sin^{2-d}\chi-1}{\cos^2\chi}\d\chi^2
-\sin^2\chi\,\d\Omega_{d-1}^2
\right]\ .
\eeq

The infinitesimal coordinate redefinition
\beq\label{SdSconformaldiff}
\chi\to\chi+m\tan\chi\left[C(\chi)+1\right]
\eeq
puts the perturbation into the conformal form
\beq\label{SdSconformalpert}
\delta q_{ij}\d x^i\d x^j
=2mC(\chi)
\left(\d\chi^2+\sin^2\chi\,\d\Omega_{d-1}^2\right)\ ,
\eeq
so that it can be compared with \eqref{3ptcorr}, \eqref{Hans},
provided that $C(\chi)$ satisfies
\beq\label{Cfirstorder}
\cos\chi\frac{\d}{\d\chi}C(\chi)+\sin\chi\,C(\chi)=-\sin^{1-d}\chi\ .
\eeq
We choose the solution that is smooth across the equator and satisfies $C(\chi)=C(\pi-\chi)$.

We next show that $C(\chi)$ satisfies a Green function equation for the same scalar operator that defines $G_1^d$, namely $D^2+d$, with delta-function sources at a pair of antipodal points. Acting on \eqref{Cfirstorder} with $\partial_\chi+(d-1)\cot\chi$ annihilates the right-hand side, and acting on the left gives
\bal
&\left[\partial_\chi+(d-1)\cot\chi\right]
\left[\cos\chi\,C'(\chi)+\sin\chi\,C(\chi)\right]\nn
&\qquad=\cos\chi\left\{
\sin^{1-d}\chi\frac{\d}{\d\chi}
\left[\sin^{d-1}\chi\,C'(\chi)\right]+dC(\chi)
\right\}=\cos\chi\,(D^2+d)C(\chi)\ .
\eal
Thus $(D^2+d)C=0$ away from the poles, with the equation at the equator following by smoothness of $C$.

To determine the delta-function coefficients, note that \eqref{Cfirstorder} gives $C'(\chi)\sim-\chi^{1-d}$ near the north pole $x_0$, where $\chi=0$. The outward flux through a small sphere $S^{d-1}_\epsilon$ at $\chi=\epsilon$ is
\beq
\int_{S^{d-1}_\epsilon}\d\Sigma\,n^iD_iC
=\operatorname{Vol}(S^{d-1})\sin^{d-1}\epsilon\,C'(\epsilon)
=-\operatorname{Vol}(S^{d-1})\ ,
\eeq
as $\epsilon\to0$, where $\d\Sigma$ is the induced measure and $n^i$ is the outward unit normal. This flux fixes the source coefficient. Similarly, the source at the other pole has the same coefficient. Therefore, 
\beq\label{Csecondorder}
(D^2+d)C(\chi)=-\operatorname{Vol}(S^{d-1})\Big[
\frac{\delta^d(x-x_0)}{\sqrt{q(x)}}
+\frac{\delta^d(x-Ax_0)}{\sqrt{q(x)}}
\Big]\ .
\eeq
This can be compared with an antipodal superposition of $G_1^{d}$ with $u=u(x,x_0)$; using  \eqref{Sgreen} and cancellation of the $\ell=1$ zero-mode gives
\beq
\frac{1}{\operatorname{Vol}(S^d)}(D^2+d)\left[G_1^{d}(u)+G_1^{d}(-u)\right]
=
\frac{\delta^{d}(x-x_0)}{\sqrt {q(x)}}
+\frac{\delta^{d}(x-Ax_0)}{\sqrt {q(x)}}\ .
\eeq
Comparing these equations shows 
\beq\label{CmatchG1}
C(\chi)=-\frac{\operatorname{Vol}(S^{d-1})}{\operatorname{Vol}(S^d)}
\left[G_1^{d}(\cos\chi)+G_1^{d}(-\cos\chi)\right]\ .
\eeq

The SdS metric perturbation given by \eqref{SdSconformalpert} and \eqref{CmatchG1} therefore gives, using \eqref{Hans}, \eqref{Hsolution}, and the normalization \eqref{Mrelat},
\beq\label{dspertH}
\delta q_{ij}(x)
=-16\pi GM \left[H_{ij}(x,x_0)+H_{ij}(x,A x_0)\right]\ , 
\eeq
where $x_0$ is the north pole and $Ax_0$ is its antipode. Equivalently, in the
canonical normalization $\tilde q_{ij}=q_{ij}+\kappa h_{ij}$,
\beq
h_{ij}(x)
=-\frac{1}{2} \kappa M\left[H_{ij}(x,x_0)+H_{ij}(x,A x_0)\right] \ .
\eeq
This  time-reflection-symmetric linearized metric perturbation precisely matches that in the  $\bra \phi \phi h \ket$ correlator \eqref{3ptcorr}, with physical mass $M\simeq \mu$ in the large mass limit, up to a factor of two associated with the commutator structure \eqref{dressdiag}.

Special cases include $d=2$, where the above solution is the Kerr-dS$_3$ geometry with $J=0$;
see \cite{Bousso:2001mw}. The geometry is locally de Sitter, with conical defects at the two poles. 
This is seen from \eqref{dSS}; defining $\gamma^2=1-2m=1-8GM$ and rescaling $t\rightarrow t/\gamma$, $r\rightarrow \gamma r$ gives 
\beq
\d s^2=-(1-r^2)\d t^2 + \frac{\d r^2}{1-r^2} + \gamma^2 r^2\d \phi^2 =-(1-r^2)\d t^2 +\frac{\d r^2}{1-r^2} +  r^2\d \varphi^2\ ,
\eeq
with $\varphi = \gamma \phi$, which is locally dS but with a different angular periodicity $\varphi \simeq \varphi + 2\pi\gamma$.  
In global coordinates, this becomes
\bal
\d s^2&=-\d t^2 + \cosh^2 t 
\left(\d\vartheta^2+\sin^2\vartheta\d\varphi^2\right) \nonumber \\
&=-\d t^2 + \cosh^2 t 
\left(\d\vartheta^2+\gamma^2\sin^2\vartheta\d\phi^2\right)\ .
\eal
Then by solving 
\beq
\frac{d\vartheta}{\sin\vartheta}=\gamma \frac{d\theta}{\sin\theta}\ ,
\eeq
the metric takes the form 
\beq
\d s^2 =-\d t^2 + \cosh^2 t \, f(\theta)^2
\left(\d\theta^2+\sin^2\theta\d\phi^2\right)\ ,
\quad f(\theta)=\gamma \frac{ \sech\left(\gamma\operatorname{arctanh}\cos\theta\right)}{\sin\theta}\ , 
\eeq
so that, to linear order,
\beq
\delta q_{ij}\d x^i\d x^j =8GM\left(\cos\theta\operatorname{arctanh}\cos\theta-1\right)
\left(\d\theta^2+\sin^2\theta\d\phi^2\right)\ .
\eeq

Also, for $d=3$, corresponding to the ordinary four-dimensional Schwarzschild-dS solution, a short calculation from \eqref{dspertH} shows that the perturbation is
\beq
\delta q_{ij}\d x^i\d x^j
=2GM\frac{ \cos 2\chi}{\sin\chi}
\big(\d\chi^2+
\sin^2\chi \, \d\Omega_2^2\big)\ .
\eeq

\section{Generalizations, discussion, and some further questions}

The previous section illustrates explicitly that, starting with the dS-invariant two-particle operator $O_2$ of \eqref{o2def}, the general dressing construction of Sec.~\ref{dresssec} gives the leading approximation to the  gauge-invariant gravitationally-dressed operator, and this, acting on the vacuum, creates the expected gravitational field.  We of course expect this to extend to more general, {\it e.g.} multiparticle, operators $O$; if these are dS-invariant (or, are made dS-invariant by the averaging prescription of \eqref{dsav}), then they can be consistently dressed, and the resulting dressed operator of \eqref{odress}
is expected to create, in the current approximation to linear order, a gravitational field sourced by the field or particle configuration created by the operator $O$.

The $\calo(\kappa^0)$ dS-invariance condition was given in \eqref{invcond},
\beq\label{invO}
\Big[\int d^dx\sqrt q  \xi^\mu_I(x) T_{n\mu}(x), O\Big] =0\ 
\eeq
where $\xi^\mu_I$ are the dS Killing vectors.
We may think of such nontrivial operators $O$ as  relational observables, where {\it e.g.} one factor of the operator is relationally defined with respect to another.  In the example of the preceding section, one may think of the first particle as relationally described with respect to the second; for example, for large mass $M$, we might imagine replacing the part of the operator creating the second particle by an operator describing a more complicated system, which for example might be a macroscopic body or other system with more appropriate properties to describe an ``observer."

This is a possible approach to giving a relational construction of an observer which is built out of the quantum fields of the existing physical system.  
This seems desirable, since all physical clocks and observers that we know are built from degrees of freedom we describe as quantum fields.
This apparently connects to the discussion of \cite{CLPW}, although there ideal observers were postulated at the beginning, and then used to describe modification of the algebra of observables, giving  von Neumann type II.  One might expect that those ideal observers furnish a description approximate to that of an actual physical observer that is made from quantum excitations of the system, though it does not seem obvious that such ideal observers exist in a truly fundamental description.  Such ideal observers have also  been considered in \cite{KoLi,CoJe}  and recently in \cite{CuKo,MNX,CSTY,HaZh}.  

Notice that while such operators satisfying the invariance condition \eqref{invO}, which can be thought of as arising from the  $\calo(\kappa^0)$ constraints, can be thought of as relational, at this $\calo(\kappa^0)$ order they have {\it no} gravitational dressing, and in this sense would not be called gravitationally dressed.  Beginning with such an operator, which we might refer to as gravitationally {\it dressable}, the corresponding operator $\hat O$ constructed in \eqref{odress} is its (leading) gravitatonally-dressed version, and arises from the condition of satisfying the constraints to $\calo(\kappa)$, or beyond for the full dressing.  This nomenclature (see also \cite{GiPe2,SGOO}) contrasts with that of \cite{CLPW}. Also, as argued in \cite{SGOO,SGcross}, the full algebraic structure of the gravitationally dressed operators appears considerably more intricate than that of a truncation that yields the type II algebras.

Such dressed operators $\hat O$ are quantum observables in the sense of being gauge invariant operators acting on the quantum states.   The question of connecting them to observations made by physical observers is one we defer for the future, and may be based on appropriate conditionals involving more complicated operators as described above.  But as quantum observables, they also have an interpretation in terms of their action on the states.  For example, we showed that the operator $\hat O_2$ acted on the vacuum state to create a state with two particles together with their gravitational field, and expect the more general operators $\hat O$ to likewise create nontrivial gravitationally-dressed excitations of the de Sitter vacuum.

This also connects to recent discussions of states of closed universes and their Hilbert spaces.  In the current description, the states $\hat O|0\rangle$ are nontrivial states for different operators $\hat O$.  In the language of a third-quantized or baby universe description\cite{Cole,GiSt,GiSt3Q}, different $\hat O$ thus create different single-universe states.  Then, the alpha-eigenstates of \cite{Cole,GiSt}, {\it e.g.} as revived in \cite{MaMa} and many subsequent references, behave as coherent multi-universe states.  We expect the different ({\it e.g.} orthogonal) single universe states to correspond to different such variables  $\alpha_I$.\footnote{Here we assume a ``free" baby universe approximation; interactions can also be considered\cite{GiSt3Q}.}  
A state with all $\alpha_I$ fixed to specified values spans a one-dimensional Hilbert space, but the higher-dimensional structure comes from the states with different values of $\alpha_I$.

In principle it appears that the construction of gravitationally dressed operators on the symmetric slice of dS can be extended to a construction of dressed operators on more general slices, though of course this involves more technical difficulty.  Study of this is left for future work.

Another related set of questions involves the careful definition of the corresponding quantum states.  At the perturbative level, in a time dependent spacetime, this involves specification of a complex structure to describe the Fock space, as recently overviewed in \cite{GiPe3}; see further references there for additional discussion.  At the level of free quantum fields, this complex structure evolves, and becomes nontrivial on more general slices.  Once 
interactions and metric perturbations are included, the story becomes more complicated\cite{ToVa2,GiPe3}.  A fuller description of this is also left for future work.

Following earlier work for flat space\cite{DoGi1} and  anti-de Sitter space\cite{GiKi}, this paper has constructed gravitationally-dressed observables to leading order in $\kappa$ in de Sitter space.  Construction of such operators to higher orders, or nonperturbatively, remains a non-trivial challenge.  This also raises very interesting questions.  We have seen that the dressed operators create the corresponding particles, together with their gravitational field.  Of course, the fully nonlinear gravitational field that we might expect from a fully-dressed observable could have significantly  non-trivial properties.  To give an example, if $\hat O$ creates one  particle plus its gravitational field, we expect  $\hat O^N$ to create $N$  particles and their gravitational field; if these particles are localized, the classical description of this field would be as a black hole.  Moreover, such black holes grow to consume the space as $N\rightarrow\infty$ -- suggesting a large modification to the algebraic structure of local quantum field theory which one might call ``algebraic spacetime disruption\cite{SGOO}."  
An important question is the structure and interpretation of the resulting algebraic structure.  We expect related statements to extend to the de Sitter context.  For example, if $\hat O_2$ is the fully-dressed version of the two particle operator considered above, the operator $\hat O_2^N$ is expected to disrupt the de Sitter spacetime for large enough $N$.  We expect the corresponding value of $N$ to be of a size associated with the discussion of the ``de Sitter locality bound" of \cite{GiMa}, which is a bound on the amount of energy that can flow through the ``neck" of de Sitter.  This gives an apparent bound on the number of nontrivial excited states of de Sitter space.

Finally, there has been considerable interest in two-dimensional toy models for gravity; it may be that some further insights may be gained through dressing constructions analogous to that of this paper, in dilaton gravity models with de Sitter spacetimes, such as Jackiw-Teitelboim gravity\cite{Jack,Teit,MTY,CJM}.

\vskip.3in
\noindent{\bf Acknowledgements} 

This material is based upon work supported in part by the U.S. Department of Energy, Office of Science, under Award Number {DE-SC}0011702, and by Heising-Simons Foundation grants \#2021-2819 and \#2024-5307.  Part of this work was finalized at the Aspen Center for Physics, which is supported by National Science Foundation grant PHY-2210452.  We thank V.~Chen, L.~Iliesiu, D.~Kolchmeyer, H.~Liu, D.~Marolf, S.~Shenker, and X. Yu for helpful conversations.

\appendix

\section{Scalar and vector harmonics on $S^d$}
\label{sphericalharmonics}

This appendix describes   conventions for and properties of scalar and vector spherical harmonics used in the main text.
All formulas are for the unit sphere $S^d$ with metric $q_{ij}$, covariant derivative $D_i$, and volume $\operatorname{Vol}(S^d)$.  We also define
\beq
u(x,x')=\cos\gamma(x,x')\ ,
\eeq
where $\gamma(x,x')$ is the geodesic distance between the two points.

\paragraph{Scalar harmonics.}
The scalar spherical harmonics $Y_{\ell a}(x)$ obey (see, {\it e.g.}, \cite{Hui:2020sch})
\bal
&D^2Y_{\ell a}(x)=-\ell(\ell+d-1)Y_{\ell a}(x)\ ,\nonumber\\
\label{SHdef}
&\ell=0,1,2,\ldots\ ,
\quad a=1,\ldots,N_\ell\ ,\quad
N_\ell=\frac{(2\ell+d-1)(\ell+d-2)!}{\ell!(d-1)!}\ ,
\eal
with the dimension $N_\ell$ given, for example, in (A.8) of \cite{Hui:2020sch} or (B.98) of \cite{Camporesi:1990wm}.

Throughout this paper, we choose a real orthonormal basis for the scalar
harmonics $Y_{\ell a}$. A real basis of scalar harmonics can be obtained
from a general complex basis by an invertible linear transformation.
Let $\widetilde Y_{\ell a}$ denote a complex basis of scalar harmonics.
We define the real harmonics by
\beq
Y_{\ell a}(x)
=\sum_b g^\ell_{ab}\widetilde Y_{\ell b}(x)\ , 
\eeq
where the coefficients $g^\ell_{ab}$ are chosen so that
$Y_{\ell a}=Y_{\ell a}^*$.  For example, the real basis may be found by working with the real/imaginary parts $(\widetilde Y_{\ell a}+\widetilde Y^*_{\ell a})/\sqrt 2$, $(\widetilde Y_{\ell a}-\widetilde Y^*_{\ell a})/(\sqrt 2 i)$ of the harmonics that are complex, together with those that are real; a nice illustrative complex basis appears in appendix A of \cite{Marolf:2008hg}.
The real spherical harmonics are normalized as
\beq\label{scalarorthonormal}
\int_{S^d}\d^d x\sqrt q\,Y_{\ell a}(x)Y_{\ell' a'}(x)
=\delta_{\ell\ell'}\delta_{aa'}\ ,
\eeq
and form a complete basis for scalar functions on $S^d$ (see \cite{Hui:2020sch}) :
\beq\label{compA}
\sum_{\ell=0}^{\infty}\sum_{a=1}^{N_\ell}
Y_{\ell a}(x)Y_{\ell a}(x')
=\frac{\delta^{d}(x-x')}{\sqrt{q(x)}}\ .
\eeq

\paragraph{Addition theorem.}
For $S^d$ with $d\geq2$, the addition theorem is\cite{WeAv}
\bal\label{additiontheorem}
\sum_{a=1}^{N_\ell}Y_{\ell a}(x)Y_{\ell a}(x')
=\frac{N_\ell}{\operatorname{Vol}(S^d)}
\frac{C_\ell^{\frac{d-1}{2}}(u)}{C_\ell^{\frac{d-1}{2}}(1)}
=\frac{1}{\operatorname{Vol}(S^d)}
\frac{2\ell+d-1}{d-1}
C_\ell^{\frac{d-1}{2}}(u)\ ,
\eal 
where $C_\ell^{\lambda}(u)$  are the Gegenbauer polynomials. The generating function for these polynomials is
\beq\label{generatingfunctions}
\frac{1}{(1-2ut+t^2)^\lambda}=\sum_{\ell=0}^{\infty}C_\ell^{\lambda}(u)t^\ell\ .
\eeq
It follows directly from this generating function that
\bal
C_\ell^{\lambda}(-u) = (-1)^\ell C_\ell^{\lambda}(u)\ .
\eal

\paragraph{Vector harmonics.}

Similar to the scalar harmonics, we choose a real orthonormal basis for vector harmonics.

For the special case of $S^2$, a convenient orthonormal basis of one-forms is given by the electric- and magnetic-type vector harmonics 
\beq
Y^E_{i,\ell a}=\frac{1}{\sqrt{\ell(\ell+1)}}D_iY_{\ell a}\ ,
\qquad
Y^B_{i,\ell a}=\frac{1}{\sqrt{\ell(\ell+1)}}\epsilon_i{}^jD_jY_{\ell a}\ ,
\qquad \ell\geq1\ .
\eeq
Here $\epsilon_{ij}$ is the Levi-Civita tensor of the unit $S^2$, normalized so that $\epsilon_{ik}\epsilon_j{}^k=q_{ij}$ and $D_i\epsilon_{jk}=0$.  With the scalar harmonics normalized as in \eqref{scalarorthonormal}, these vector harmonics obey (see, {\it e.g.}  \cite{Hui:2020sch})
\beq
\int_{S^2}\d^2x\sqrt q\,q^{ij}Y^A_{i,\ell a}(x)Y^B_{j,\ell' a'}(x)
=\delta^{AB}\delta_{\ell\ell'}\delta_{aa'}\ ,
\qquad A,B=E,B\ .
\eeq
The $E$-modes are longitudinal, while the $B$-modes are transverse.  Both are eigenvectors of the  Laplacian $D^2=D^i D_i$ on one-forms:
\beq
D^2Y^E_{i,\ell a}=\left[-\ell(\ell+1)+1\right]Y^E_{i,\ell a}\ ,
\qquad
D^2Y^B_{i,\ell a}=\left[-\ell(\ell+1)+1\right]Y^B_{i,\ell a}\ .
\eeq

More generally on $S^d$, vector harmonics decompose into longitudinal and transverse families \cite{Hui:2020sch, Camporesi:1990wm, Loganayagam:2025jmw}. The normalized longitudinal modes are
\beq
Y^L_{i,\ell a}=\frac{1}{\sqrt{\ell(\ell+d-1)}}D_iY_{\ell a}\ ,
\quad \ell\geq1\ ,\quad a=1,\ldots,N_\ell\ ,\quad
\eeq
with $N_\ell$ given in \eqref{SHdef}, and obey
\beq
D^2Y^L_{i,\ell a}=\left[-\ell(\ell+d-1)+(d-1)\right]Y^L_{i,\ell a}\ .
\eeq
The transverse modes $Y^T_{i,\ell \ta}$ are labelled by $\ell$ and $\ta$ with
\bal
\ell\geq1\ , \quad \ta=1,\ldots, N_\ell^T\ ,\quad
N_\ell^T=\frac{\ell(\ell+d-1)(2\ell+d-1)(\ell+d-3)!}{(d-2)!(\ell+1)!}\ ,
\eal
and satisfy
\beq
D^iY^T_{i,\ell \ta}=0\ ,
\qquad
D^2Y^T_{i,\ell \ta}=\left[-\ell(\ell+d-1)+1\right]Y^T_{i,\ell \ta}\ .
\eeq
For $\ell=1$, these transverse modes have degeneracy $N_1^T=d(d+1)/2$ and coincide with the Killing one-forms on $S^d$. The index $\ta$ denotes the appropriate degeneracy in each family. Together, the longitudinal and transverse harmonics give the completeness relation on one-forms,
\bal\label{vectorcomplete}
\sum_{\ell=1}^{\infty}\sum_a
Y^L_{i,\ell a}(x)Y^{Lj}_{\ell a}(x')
+
\sum_{\ell=1}^{\infty}\sum_\ta
Y^T_{i,\ell \ta}(x)Y^{Tj}_{\ell \ta}(x')
=\delta_i^j\,\frac{\delta^{d}(x-x')}{\sqrt{q(x)}}\ .
\eal
where  $\delta_i^j$ is the identity matrix on the tangent space.

\section{Green functions on $S^d$}
\label{gfapp}

This appendix solves for Green functions used in the main text.
For $d\geq2$ and a non-negative integer $s$, define a Green function problem 
\beq\label{GFeqn}
\frac{1}{\operatorname{Vol}(S^d)}
\left[D^2+s(s+d-1)\right]G_s^{d}(x,x')
=
\frac{\delta^{d}(x-x')}{\sqrt{q(x)}}
-
\sum_{a=1}^{N_s}Y_{sa}(x)Y_{sa}(x')\ .
\eeq
Solutions can be found in the form $G_s^d(x,x')= G_s^{d}(u)$, with $u=u(x,x')$. The second term on the right is necessary to project out zero modes $Y_{sa}$ of $D^2+s(s+d-1)$.
We fix the solution $G_s^{d}$ by requiring it  to have no $\ell=s$ component. To find the solutions, begin with the Ansatz 
\beq
G_s^{d}(u) = \sum_{\ell=0,\neq s}^\infty c_\ell C_\ell^{\frac{d-1}{2}}(u)\ .
\eeq
The expansion \eqref{additiontheorem} in spherical harmonics then shows $G_s^{d}(u)$ obeys \eqref{GFeqn} with the coefficients
\beq\label{Gsmodes}
G_s^{d}(u)=
\sum_{\ell=0,\,\neq s}^{\infty}
\frac{1}{(s-\ell)(\ell+d+s-1)}\frac{2\ell+d-1}{d-1} C_\ell^{\frac{d-1}{2}}(u)\ .
\eeq

Such sums can be evaluated using the generating function \eqref{generatingfunctions}.  To begin, acting on it with
$\left(1+\frac{2t}{d-1}\partial_t\right)$ gives the formula
\bal\label{gegenbauerpoisson}
\sum_{\ell=0}^{\infty}t^\ell \frac{2\ell+d-1}{d-1}
C_\ell^{\frac{d-1}{2}}(u)
&=\left(1+\frac{2t}{d-1}\partial_t\right)\sum_{\ell=0}^{\infty}t^\ell C_\ell^{\frac{d-1}{2}}(u) \nonumber \\
&=\left(1+\frac{2t}{d-1}\partial_t\right)
(1-2ut+t^2)^{-\frac{d-1}{2}}
=
\frac{1-t^2}{(1-2ut+t^2)^{\frac{d+1}{2}}}\ .
\eal

For the case $s=0$, the additional denominator in \eqref{Gsmodes} can be rewritten as
\beq
-\frac{1}{\ell(\ell+d-1)}
=
\frac{1}{d-1}\left(\frac{1}{\ell+d-1} - \frac{1}{\ell}\right)
=
\frac{1}{d-1}\int_0^1 \mathrm dt\,
\left(t^{d-2}-\frac{1}{t}\right)t^\ell\ ,
\qquad \ell\geq1\ .
\eeq
Using this in \eqref{gegenbauerpoisson}, and using $C_0^\lambda(u)=1$, then gives the integral representation
\beq\label{G0integral}
G_0^{d}(u)=
\frac{1}{d-1}\int_0^1 \mathrm dt\,
\left(t^{d-2}-\frac{1}{t}\right)
\left[
\frac{1-t^2}{(1-2ut+t^2)^{\frac{d+1}{2}}}-1
\right]\ .
\eeq

For the case $s=1$, the $\ell=0$ term in \eqref{Gsmodes} contributes $1/d$ and the $\ell=1$ term is omitted.  For the remaining terms,
\beq
\frac{1}{(1-\ell)(\ell+d)}
=\frac{1}{d+1}\left(\frac{1}{\ell+d} - \frac{1}{\ell-1}\right)=
\frac{1}{d+1}\int_0^1 \mathrm dt\,
\left(t^{d-1}-\frac{1}{t^2}\right)t^\ell\ ,
\qquad \ell\geq2\ .
\eeq
Combining this with \eqref{gegenbauerpoisson} and using $C_1^\lambda(u)=2\lambda u$ then gives the integral representation
\beq\label{G1integral}
G_1^{d}(u)
=
\frac{1}{d}
+
\frac{1}{d+1}\int_0^1 \mathrm dt\,
\left(t^{d-1}-\frac{1}{t^2}\right)
\left[
\frac{1-t^2}{(1-2ut+t^2)^{\frac{d+1}{2}}}
-1-(d+1)ut
\right]\ ;
\eeq
this Green function was also recently used in \cite{MNX}.

The integrals \eqref{G0integral} and \eqref{G1integral} can be evaluated explicitly for each dimension $d$, yielding the following examples.  To simplify the expressions for odd dimensions, define
\beq
A(u)=\pi-\arccos u\ .
\eeq
The results are
\beq
\begin{aligned}
G_0^{2}(u)&=1+\log \frac{1-u}{2}\ ,\\
G_0^{3}(u)&=\frac{1}{4}-\frac{A(u)u}{2\sqrt{1-u^2}}\ ,\\
G_0^{4}(u)&=\frac{4-7u}{9(1-u)}+\frac{1}{3}\log \frac{1-u}{2}\ ,\\
G_0^{5}(u)&=\frac{3-5u^2}{16(1-u^2)}
-\frac{A(u)u(3-2u^2)}{8(1-u^2)^{3/2}}\ ,\\
G_0^{6}(u)&=\frac{23-71u+43u^2}{75(1-u)^2}
+\frac{1}{5}\log \frac{1-u}{2}\ ,\\
G_0^{7}(u)&=\frac{22-71u^2+40u^4}{144(1-u^2)^2}
+\frac{A(u)u(-15+20u^2-8u^4)}{48(1-u^2)^{5/2}}\ ,\\
G_0^{8}(u)&=\frac{176-759u+906u^2-337u^3}{735(1-u)^3}
+\frac{1}{7}\log \frac{1-u}{2}\ ,\\
G_0^{9}(u)&=\frac{50-237u^2+266u^4-94u^6}{384(1-u^2)^3}
+\frac{A(u)u(-35+70u^2-56u^4+16u^6)}{128(1-u^2)^{7/2}}\ ,\\
G_0^{10}(u)&=\frac{563-3089u+5466u^2-4049u^3+1091u^4}{2835(1-u)^4}
+\frac{1}{9}\log \frac{1-u}{2}\ ,
\end{aligned}
\eeq
and
\beq
\begin{aligned}
G_1^{2}(u)&=1+\frac{4}{3}u+u\log \frac{1-u}{2}\ ,\\
G_1^{3}(u)&=\frac{u}{4}+\frac{A(u)(1-2u^2)}{2\sqrt{1-u^2}}\ ,\\
G_1^{4}(u)&=\frac{10+13u-28u^2}{15(1-u)}
+u\log \frac{1-u}{2}\ ,\\
G_1^{5}(u)&=\frac{u(13-16u^2)}{24(1-u^2)}
+\frac{A(u)(3-12u^2+8u^4)}{8(1-u^2)^{3/2}}\ ,\\
G_1^{6}(u)&=\frac{56+64u-359u^2+232u^3}{105(1-u)^2}
+u\log \frac{1-u}{2}\ ,\\
G_1^{7}(u)&=\frac{u(35-84u^2+46u^4)}{48(1-u^2)^2}
+\frac{A(u)(5-30u^2+40u^4-16u^6)}{16(1-u^2)^{5/2}}\ ,\\
G_1^{8}(u)&=\frac{144+131u-1518u^2+2013u^3-776u^4}{315(1-u)^3}
+u\log \frac{1-u}{2}\ ,\\
G_1^{9}(u)&=\frac{u(1667-5926u^2+6456u^4-2272u^6)}{1920(1-u^2)^3}
+\frac{A(u)(35-280u^2+560u^4-448u^6+128u^8)}{128(1-u^2)^{7/2}}\ ,\\
G_1^{10}(u)&=\frac{1408+876u-21159u^2+43096u^3-33479u^4+9236u^5}{3465(1-u)^4}
+u\log \frac{1-u}{2}\ .
\end{aligned}
\eeq
Related closed-form expressions are tabulated in Tables 1 and 2 of \cite{IglewskaNowak:2023wavelet}.

\mciteSetMidEndSepPunct{}{\ifmciteBstWouldAddEndPunct.\else\fi}{\relax}
\bibliographystyle{utphys}
\bibliography{ds-dress}{}

\end{document}